\documentclass[oneside,showpacs,twocolumn,superscriptaddress,pra]{revtex4}%
\usepackage{amsmath}
\usepackage{amsfonts}
\usepackage{amssymb}
\usepackage{graphicx}%
\begin{document}
\title{Spatial Localization and Relativistic Transformation of Quantum Spins}
\author{Hui Li}
\email{lhuy@mail.ustc.edu.cn}
\affiliation{Hefei National Laboratory for
Physical Sciences at Microscale \& Department of Modern Physics,
University of Science and Technology of China, Hefei, Anhui
230026, P.R. China}
\author{Jiangfeng Du}
\email{djf@ustc.edu.cn}
\affiliation{Hefei National Laboratory for
Physical Sciences at Microscale \& Department of Modern Physics,
University of Science and Technology of China, Hefei, Anhui
230026, P.R. China}
\affiliation{Department of Physics, National
University of Singapore, Lower Kent Ridge, Singapore 119260,
Singapore}

\begin{abstract}
The purity of the reduced state for spins which is pure in the
rest frame will most likely appear to degrade because spin and
momentum become mixed when viewed by a moving observer. We show
that such boost-induced decrease in spin purity observed in a
moving reference frame is intrinsically related to the spatial
localization properties of the wave package observed in the rest
frame. Furthermore, we prove that, for any localized pure state
with separable spin and momentum in the rest frame, its reduced
density matrix for spins inevitably appears to be mixed whenever
viewed from a moving reference frame.

\end{abstract}
\maketitle

\section{Introduction}

One of the most nontrivial and striking observations of relativistic
thermodynamics \cite{b1} is that probability distributions can depend on
frames. Consequently, entropy and information may change if viewed from
different reference frames \cite{r0}. Recently, relativistic quantum
information theory has attracted particular interests
\cite{r1,r2,r3,r4,r5,r6,r7}. Current investigations show that a single quantum
spin is not covariant under Lorentz transformations \cite{r1}, and maximal
entanglement between spins in the rest frame will most likely degrade due to
mixing with the momentum if viewed from a moving frame \cite{r2,r3}, depending
on the initial momentum wave function. For many quantum information protocols
\cite{b5}, coherence and entanglement are extremely important and expensive.
The observation that coherence and entanglement of the reduced state for spins
may degrade when viewed in moving frames implies that there are particular
problems for relativistic quantum information processing, particularly for
relativistic quantum communication.

It is known that, for a single quantum spin, if and only if we consider the
momentum eigenstates (plane waves), can the reduced density matrix for the
spin be covariant under Lorentz transformations. But momentum eigenstates are
not localized so they may be difficult in feasible applications \cite{r1}. The
similar difficulties exist for multipartite states \cite{ft1}.

In this paper, we investigate the Lorentz boost-induced decrease
in the purity of the reduced density matrix for spins, when a
state which has pure reduced state for spins in the rest frame is
viewed from a moving reference frame. Taken to the leading order,
the decrease in spin purity observed in the moving frame is linear
with respect to the momentum mean square deviation observed in the
rest frame, which according to the position-momentum uncertainty
relationship can be reasonably regarded as a measure of how much
the spacial wave package is localized. We also present numerical
studies as instance of our general analysis. Furthermore, we prove
that, for any localized pure state with separable spin and
momentum in the rest frame, its reduced density matrix for spins
cannot be covariant under any Lorentz boosts, i.e. it inevitably
appears to be mixed when viewed from a moving reference frame.
Considering that in practical applications states should be
localized, our results may have important consequences for
relativistic quantum information processing.

\section{General Relationship between the Boost-induced Spin
Depurification and the Spatial Localization}

\subsection{A Simple Example: Single Spin-$1/2$ Massive Particle}

We start by briefly recalling Peres et al.'s paper \cite{r1}. Consider a spin
half massive particle (of mass $m$) that is prepared with spin in the $z$
direction. The spin state can be represented by the Bloch vector
$\mathbf{n}=(n_{x},n_{y},n_{z})$ with $n_{x}=n_{y}=0$ and $n_{z}=1$. The
momentum wave function is a Gaussian $g(\mathbf{p})\propto\exp(-\mathbf{p}%
^{2}/2w^{2})$. When viewed by an observer moving in the $x$ direction, the
Lorentz-transformed Bloch vector is $\mathbf{n}^{\prime}=(n_{x}^{\prime}%
,n_{y}^{\prime},n_{z}^{\prime})$ with $n_{x}^{\prime}=n_{y}^{\prime}=0$ yet
$n_{z}^{\prime}<1$. It is shown in Ref. \cite{r1} that $1-n_{z}^{\prime
}\propto w^{2}$ to the leading order of $w/m\ll1$. By denoting the
Lorentz-transformed density matrix for spin as $\rho^{\prime}$, the Lorentz
boost-induced decrease in its purity is $1-\operatorname{tr}(\rho^{\prime
2})\propto w^{2}$. Meanwhile, for this particular case the momentum mean
square deviation is $\langle\Delta p_{\mu}^{2}\rangle\propto w^{2}$
($\mu=x,y,z$), hence $1-\operatorname{tr}(\rho^{\prime2})\propto\langle\Delta
p_{\mu}^{2}\rangle$. According to the position-momentum uncertainty
relationship $\langle\Delta x_{\mu}^{2}\rangle\langle\Delta p_{\mu}^{2}%
\rangle\geqslant\hbar^{2}/4$, the smaller $\langle\Delta x_{\mu}^{2}\rangle$
is, the larger $\langle\Delta p_{\mu}^{2}\rangle$ is and so is
$1-\operatorname{tr}(\rho^{\prime2})$. This suggests that the more the wave
package is localized in space, the more the boost-induced decrease in spin
purity is when viewed in a moving frame.

\subsection{Generalization to Multiple Massive Particles with Arbitrary Spins}

The above observation can indeed generalize to states of multiple
massive particles with arbitrary spin quantum number. Consider a
pure quantum state with separable spin and momentum in the rest
frame. The system consists of $N$ massive particles, labelled by
$k=1,\cdots,N$. The spin quantum number of particle $k$ is
$s_{k}$, and mass $m_{k}>0$. The reduced density matrix for spins
viewed from the rest frame is denoted by $\rho$, which is pure
with $\operatorname{tr}(\rho^{2})=1$. The normalized momentum wave
function in the
rest frame is denote by $g(\mathcal{P})$, where $\mathcal{P}:=(p_{1x}%
,p_{1y},p_{1z},\cdots)$ for compactness of notation. To a moving observer
Lorentz transformed by $\Lambda^{-1}$, the state appears to be transformed by
$\Lambda^{\otimes N}$. Because the Lorentz-transformed state viewed in the
moving frame differs from rest-frame one by unitary transformations, the
purity will not change provided we do not trace out a part of the state.
However, in looking at spins, tracing out over the momentum degrees of freedom
is implied. To the Lorentz-transformed observer, the spin and momentum may
appear to be entangled, thus the purity of spins may appear to degrade viewed
by this observer.

The reduced density matrix for spins viewed by the moving observer is
\cite{r1,r2,r3}%
\begin{equation}
\rho^{\prime}=\int\left\vert g(\mathcal{P})\right\vert ^{2}U_{\Lambda
}(\mathcal{P})\rho U_{\Lambda}^{\dag}(\mathcal{P})\widetilde{\mbox{d}}%
\mathcal{P},\label{eq 1}%
\end{equation}
where, for compactness, we define $U_{\Lambda}(\mathcal{P}):=U^{s_{1}}%
(\Lambda,\mathbf{p}_{1})\otimes\cdots\otimes U^{s_{N}}(\Lambda,\mathbf{p}%
_{N})$ with $U^{s}(\Lambda,\mathbf{p})$ being the spin-$s$ representation of
the Wigner rotation $R(\Lambda,\mathbf{p})$ \cite{b3}. $\widetilde
{\mbox{d}}\mathcal{P}:=\widetilde{\mbox{d}}\mathbf{p}_{1}\cdots\widetilde
{\mbox{d}}\mathbf{p}_{N}$ and $\widetilde{\mbox{d}}\mathbf{p}_{k}\ $is the
Lorentz invariant integration measure (defined in Ref. \cite{b3}). We
represent the Lorentz-transformed spin purity by $\operatorname{tr}%
(\rho^{\prime2})$, which can be calculated by%
\begin{equation}
\operatorname{tr}(\rho^{\prime2})=\iint\left\vert g(\mathcal{P})\right\vert
^{2}\left\vert g(\mathcal{P}^{\prime})\right\vert ^{2}\Gamma_{\rho}^{\Lambda
}(\mathcal{P},\mathcal{P}^{\prime})\widetilde{\mbox{d}}\mathcal{P}%
\hspace{\stretch{4}}\widetilde{\mbox{d}}\mathcal{P}^{\prime}.\label{eq 2}%
\end{equation}
where%
\begin{equation}
\Gamma_{\rho}^{\Lambda}(\mathcal{P},\mathcal{P}^{\prime}):=\operatorname{tr}%
[U_{\Lambda}(\mathcal{P})\rho U_{\Lambda}^{\dag}(\mathcal{P})U_{\Lambda
}(\mathcal{P}^{\prime})\rho U_{\Lambda}^{\dag}(\mathcal{P}^{\prime
})].\label{eq 2-2}%
\end{equation}
Denoting $\langle\mathbf{\cdot}\rangle$ as the mean value (observed in the
rest frame), we define $\Delta p_{k\mu}^{(\prime)}:=p_{k\mu}^{(\prime
)}-\langle p_{k\mu}^{(\prime)}\rangle$ ($\mu=x,y,z$) and $\Delta
\mathcal{P}^{(\prime)}=\mathcal{P}^{(\prime)}-\langle\mathcal{P}^{(\prime
)}\rangle$. Then $\Gamma_{\rho}^{\Lambda}(\mathcal{P},\mathcal{P}^{\prime})$
can be expanded into power series with respect to $\Delta\mathcal{P}$ and
$\Delta\mathcal{P}^{\prime}$, noting that $\Gamma_{\rho}^{\Lambda}%
(\mathcal{P},\mathcal{P}^{\prime})=\Gamma_{\rho}^{\Lambda}(\mathcal{P}%
^{\prime},\mathcal{P})$:%
\begin{equation}
\Gamma_{\rho}^{\Lambda}(\mathcal{P},\mathcal{P}^{\prime})=1-\frac{1}{2}%
(\Delta\mathcal{P},\Delta\mathcal{P}^{\prime})\left(
\begin{array}
[c]{cc}%
\mathcal{U} & \mathcal{V}^{T}\\
\mathcal{V} & \mathcal{U}%
\end{array}
\right)  \left(
\begin{array}
[c]{c}%
\Delta\mathcal{P}^{T}\\
\Delta\mathcal{P}^{\prime T}%
\end{array}
\right)  +\cdots,\label{eq 3}%
\end{equation}
where $3N$-dimensional matrices $\mathcal{U}$ and $\mathcal{V}$ are real
functions of $\Lambda$ and $\rho$. When $\Delta p_{k\mu}=\Delta p_{k\mu
}^{\prime}=0$, $\Gamma_{\rho}^{\Lambda}(\mathcal{P},\mathcal{P}^{\prime})$
reaches its maximum, i.e. $\Gamma_{\rho}^{\Lambda}(\langle\mathcal{P\rangle
},\langle\mathcal{P\rangle})=1$. So in the r.h.s. of Eq. (\ref{eq 3}), the
zero-order term is $1$, the first-order terms vanish, and $\mathcal{U}$ is
positive-semidefinite. Terms of higher than second orders are not explicitly
presented here. Using $\langle\Delta p_{k\mu}\rangle=\langle\Delta p_{k\mu
}^{\prime}\rangle=0$ and $\langle\Delta\mathcal{P}\cdot\mathcal{U}\cdot
\Delta\mathcal{P}^{T}\rangle=\langle\Delta\mathcal{P}^{\prime}\cdot
\mathcal{U}\cdot\Delta\mathcal{P}^{\prime T}\rangle$, we see that the
boost-induced decrease in spin purity, to the leading order, is%
\begin{equation}
1-\operatorname{tr}(\rho^{\prime2})\simeq\langle\Delta\mathcal{P}%
\cdot\mathcal{U}\cdot\Delta\mathcal{P}^{T}\rangle.
\end{equation}
The matrix $\mathcal{U}$ can be diagonalized as $\mathcal{U}=\mathcal{M}%
\cdot\mathcal{D}\cdot\mathcal{M}^{T}$, where $\mathcal{M}$ is real and
orthogonal, $\mathcal{D}=\operatorname*{diag}(\mathcal{D}_{1},\cdots
,\mathcal{D}_{3N})$ with $\mathcal{D}_{\lambda}\geqslant0$ ($\lambda
=1,\cdots,3N$) due to that $\mathcal{U}$ is positive-semidefinite. By denoting
$\mathcal{Q}=\mathcal{P}\cdot\mathcal{M}$ and $\Delta\mathcal{Q}%
=\mathcal{Q}-\langle\mathcal{Q}\rangle=\Delta\mathcal{P\cdot M}$, we have%
\begin{equation}
1-\operatorname{tr}(\rho^{\prime2})\simeq\sum_{\lambda=1}^{3N}\mathcal{D}%
_{\lambda}\langle\Delta\mathcal{Q}_{\lambda}^{2}\rangle.\label{eq 4}%
\end{equation}
Inspired by $[\widehat{x}_{k\mu},\widehat{p}_{k^{\prime}\mu^{\prime}}%
]=i\hbar\delta_{kk^{\prime}}\delta_{\mu\mu^{\prime}}$, we define
$\mathcal{X}=(x_{1x},x_{1y},x_{1z},\cdots)\cdot\mathcal{M}$ so that
$[\mathcal{\widehat{X}}_{\lambda},\mathcal{\widehat{Q}}_{\lambda^{\prime}%
}]=i\hbar\delta_{\lambda\lambda^{\prime}}$, leading to the uncertainty
relationship $\langle\Delta\mathcal{X}_{\lambda}^{2}\rangle\langle
\Delta\mathcal{Q}_{\lambda}^{2}\rangle\geqslant\hbar^{2}/4$. Hence from Eq.
(\ref{eq 4}) we obtain%
\begin{equation}
\operatorname{tr}(\rho^{\prime2})\alt1-\frac{\hbar^{2}}{4}\sum_{\lambda
=1}^{3N}\frac{\mathcal{D}_{\lambda}}{\langle\Delta\mathcal{X}_{\lambda}%
^{2}\rangle}.\label{eq 6}%
\end{equation}

Here we shall note that both $\langle\Delta\mathcal{Q}_{\lambda}^{2}\rangle$
and $\langle\Delta\mathcal{X}_{\lambda}^{2}\rangle$ are observed in the rest
frame, while $\operatorname{tr}(\rho^{\prime2})$ is the purity observed in the
moving frame. In addition, $\mathcal{D}$ and $\mathcal{M}$ are functions of
$\Lambda$ and $\rho$. For any $\rho$, when $\Lambda$ degenerate to a pure
three-dimensional rotation, we can see in Eq. (\ref{eq 2-2}) that
$\Gamma_{\rho}^{\Lambda}(\mathcal{P},\mathcal{P}^{\prime})=1$ and then in Eq.
(\ref{eq 4}) that $\mathcal{D}_{\lambda}=0$ for all $\lambda$, hence obviously
$\operatorname{tr}(\rho^{\prime2})=1$. Equation (\ref{eq 6}) also indicates
that the purity of the Lorentz-transformed reduced spin state is (to the
leading order) bounded by its spatial localization properties.

Equations (\ref{eq 4}) and (\ref{eq 6}) dictate the intrinsic relation between
the boost-induced decrease in spin purity (observed in the moving frame) and
the spatial localization of the wave package (observed in the rest frame).
Indeed, $\mathcal{X}_{\lambda}$ is a linear combination of $(x_{1x}%
,x_{1y},x_{1z},\cdots)$. The more the wave package is localized (the less
$\langle\Delta\mathcal{X}_{\lambda}^{2}\rangle$ is), the more mixed the
reduced state for spins becomes when viewed in a moving reference frame.
According to the uncertainty relationship $\langle\Delta\mathcal{X}_{\lambda
}^{2}\rangle\langle\Delta\mathcal{Q}_{\lambda}^{2}\rangle\geqslant\hbar^{2}/4
$, $\sum_{\lambda=1}^{3N}\mathcal{D}_{\lambda}\langle\Delta\mathcal{Q}%
_{\lambda}^{2}\rangle$ can be reasonably regarded as a measure of how much the
wave package is localized in space.

We shall note that $\mathcal{D}$ and $\mathcal{M}$ do not explicitly depend on
the momentum wave function (the implicit dependence is through the mean value
$\langle\mathcal{P}\rangle$ because $\Delta\mathcal{P}=\mathcal{P}%
-\langle\mathcal{P}\rangle$). When the state is not strongly localized,
whatever the momentum wave function as long as the localization is the same
(i.e. $\sum_{\lambda=1}^{3N}\mathcal{D}_{\lambda}\langle\Delta\mathcal{Q}%
_{\lambda}^{2}\rangle$ is the same), the same reduced state for
spins suffers the same amount of boost-induced decrease in spin
purity when viewed from the moving frame. This provides a general
and feasible method to possibly estimate how mixed the reduced
spin state would appear by the spatial localization properties,
which might be useful in practical relativistic quantum
information processing. In addition, for multipartite states whose
spins are not entangled, position (momentum) coordinates of
different particles will not be mixed in $\mathcal{X}_{\lambda}$
($\mathcal{Q}_{\lambda}$). The r.h.s. of Eq. (\ref{eq 4}), as well
as that of Eq. (\ref{eq 6}), turns out to be a sum over each
individual particles. While interestingly, if the spins are
entangled, position (momentum) coordinates of different particles
will in general be mixed in $\mathcal{X}_{\lambda}$
($\mathcal{Q}_{\lambda}$). This implies that, for the present
case, in measuring how much the spatial wave package is localized,
the correlation (entanglement) between the spins needs also to be
taken into account.

\subsection{A Further Illustration: Two Spin-$1/2$ Massive Particles}

\begin{figure}[b]
\includegraphics[width=\columnwidth]{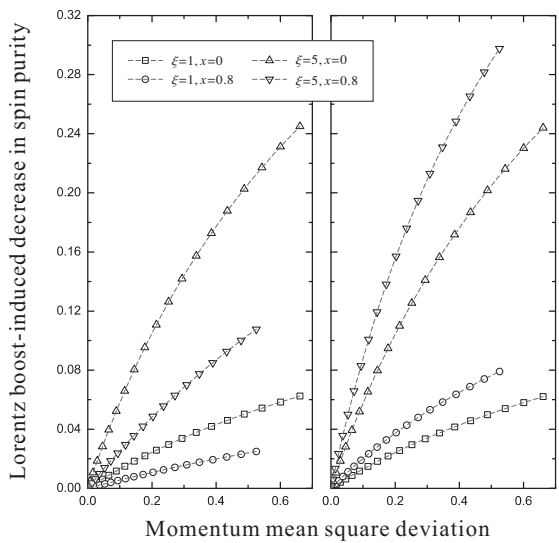}\caption{Lorentz boost-induced
decrease of spin purity (observed in the moving frame) versus the
momentum mean square deviation (observed in the rest frame), for
two states, $|\psi ^{-}\rangle$ (at left) and $|\psi^{+}\rangle$
(at right), with momentum wave function in Eq. (\ref{eq 5}). For
each set of $(\xi,x)$, $\sigma/m$ takes
value from $0.025$ to $0.5$ with a step $0.025$. Here we adopt the natural units:
$c=\hbar=1$, and $m=1$ so that the quantities are all dimensionless.}%
\label{figure}%
\end{figure}

As an illustrative example, we numerically study the case of two spin half
particles (of mass $m$) with momentum wave function being the
\textquotedblleft entangled Gaussian\textquotedblright\ as presented in Ref.
\cite{r3}:%
\begin{align}
&  g\left(  \mathbf{p}_{1},\mathbf{p}_{2}\right)  \nonumber\\
&  =\sqrt{\frac{1}{\mathsf{N}}\exp\left[  -\frac{\mathbf{p}_{1}^{2}%
+\mathbf{p}_{2}^{2}}{4\sigma^{2}}\right]  \exp\left[  -\frac{\mathbf{p}%
_{1}^{2}+\mathbf{p}_{2}^{2}-2x\mathbf{p}_{1}\cdot\mathbf{p}_{2}}{4\sigma
^{2}\left(  1-x^{2}\right)  }\right]  },\label{eq 5}%
\end{align}
where $\mathsf{N}$ is the normalization, $\sigma\geqslant0$ is the
\textquotedblleft width\textquotedblright\ and $x\in\left[  0,1\right)  $. The
Lorentz transformation is chosen to be a pure boost $L(\boldsymbol{\xi})$ in
the $z$ direction, where $\boldsymbol{\xi}$ is the rapidity and denote
$\xi=|\boldsymbol{\xi}|$, as defined in Ref. \cite{r3}. For this particular
momentum wave function, $\langle\Delta p_{k\mu}^{2}\rangle$ is the same for
any $k$ and $\mu$. Moreover, when $\langle\Delta p_{k\mu}^{2}\rangle$ and
$\langle\Delta p_{k^{\prime}\mu^{\prime}}^{2}\rangle$ are relatively small,
$\langle\Delta p_{k\mu}\Delta p_{k^{\prime}\mu^{\prime}}\rangle$ is (if not
zero) proportional to $[\langle\Delta p_{k\mu}^{2}\rangle\langle\Delta
p_{k^{\prime}\mu^{\prime}}^{2}\rangle]^{1/2}$, with the proportion depending
upon $x$. Thus for this particular case, Eq. (\ref{eq 4}) reduces to
$1-\operatorname{tr}(\rho^{\prime2})\propto\langle\Delta\mathbf{p}_{1}%
^{2}\rangle$. Figure \ref{figure} depicts the relation between
$1-\operatorname{tr}(\rho^{\prime2})$ and $\langle\Delta\mathbf{p}_{1}%
^{2}\rangle$, for the spin parts being $|\psi^{-}\rangle$ and
$|\psi ^{+}\rangle$. When the momentum mean square deviation is
relatively small, the relation can be well described by linearity,
confirming the validity of Eq. (\ref{eq 4}). As the momentum mean
square deviation increases, corresponding to stronger
localization, the boost-induced decrease in spin purity increases
monotonously. The deviation from linearity is due to terms of
higher than second orders, such as $\langle\Delta
p_{k\mu}^{3}\rangle$ and $\langle\Delta p_{k\mu}^{4}\rangle$ etc.
However, such terms can also be regarded in some sense as measures
of the spatial localization of the wave package.

\subsection{A Theorem}

In the remaining part of this paper, we prove the following theorem.

\textbf{Theorem:} \textit{When a pure state with separable spin
and momentum in the rest frame is viewed from a moving reference
frame, its reduced density matrix for spins necessarily appears to
be mixed if its spatial wave package is localized.}

Here we shall first specify the meaning of being localized. We regard a state
localized in the sense that its position wave function could be
\textit{commonly normalized}, i.e. it is square-integrable. This requirement
excludes nonlocalized states, such as momentum eigenstates and the singular
ones presented in Ref. \cite{r3}. Since momentum wave function is the Fourier
transformation of position wave function, being localized means that, in our
case, $g(\mathcal{P})$ is square-integrable, i.e. $G(\mathcal{P}%
):=|g(\mathcal{P})|^{2}\widetilde{\mbox{d}}\mathcal{P}/\mbox{d}\mathcal{P}%
\geqslant0$ is \textit{integrable} on $\mathbb{R}^{3N}$ in the sense of
\textit{Lebesgue integration}, and it is normalized as $\int G(\mathcal{P}%
)\mbox{d}\mathcal{P}=\int_{\operatorname*{supp}(G(\mathcal{P}))}%
G(\mathcal{P})\mbox{d}\mathcal{P}=1$, where $\operatorname*{supp}%
(G(\mathcal{P})):=\{\mathcal{P}\mid G(\mathcal{P})>0\}$ is the support of
$G(\mathcal{P})$.

We denote $\mathcal{K}=(\mathcal{P},\mathcal{P}^{\prime})\in\mathbb{R}^{6N} $,
$\mathcal{T}(\mathcal{K})=\Gamma_{\rho}^{\Lambda}(\mathcal{P},\mathcal{P}%
^{\prime})\in\lbrack0,1]$, and $\mathcal{G}(\mathcal{K})=G(\mathcal{P}%
)G(\mathcal{P}^{\prime})\geqslant0$ for compactness. Let $\Omega
_{g}=\operatorname*{supp}(\mathcal{G}(\mathcal{K}))=\{\mathcal{K}%
\mid\mathcal{G}(\mathcal{K})>0\}=\operatorname*{supp}(G(\mathcal{P}%
))\times\operatorname*{supp}(G(\mathcal{P}^{\prime}))$, $\Omega_{t}%
=\{\mathcal{K}\mid\mathcal{T}(\mathcal{K})=1\}$, and $m(\cdot)$ be the
Lebesgue measure in $\mathbb{R}^{6N}$. The integral in Eq. (\ref{eq 2}) now
turns to be the Lebesgue integral over $\Omega_{g}$: $\operatorname{tr}%
(\rho^{\prime2})=\int_{\Omega_{g}}\mathcal{G}(\mathcal{K})\mathcal{T}%
(\mathcal{K})\mbox{d}\mathcal{K}.$

Since the state is localized, $\mathcal{G}(\mathcal{K})$ must be integrable.
In addition, $\int_{\Omega_{g}}\mathcal{G}(\mathcal{K})\mbox{d}\mathcal{K}%
=[\int_{\operatorname*{supp}(G(\mathcal{P}))}G(\mathcal{P})\mbox{d}\mathcal{P}%
]^{2}=1$ implies that $\mathcal{G}(\mathcal{K})$ is bounded almost everywhere.
Hence one must have $m(\Omega_{g})>0$ \cite{b4}. Otherwise supposing
$m(\Omega_{g})=0$, one necessarily encounters the contradiction that
$\int_{\Omega_{g}}\mathcal{G}(\mathcal{K})\mbox{d}\mathcal{K}=0$ and
$\int_{\operatorname*{supp}(G(\mathcal{P}))}G(\mathcal{P})\mbox{d}\mathcal{P}%
=0$.

Denote $x=(\mathbf{p}_{2},\cdots,\mathbf{p}_{N},\mathbf{p}_{1}^{\prime}%
,\cdots,\mathbf{p}_{N}^{\prime})$ and $\Omega_{t}(x)=\{\mathbf{p}_{1}%
\mid(\mathbf{p}_{1},x)\in\Omega_{t}\}$. It can be easily verified that
$\Omega_{t}(x)\ $has Lebesgue measure zero in $\mathbb{R}^{3}$ for any given
$x$ when $\Lambda$ does not degenerate to a pure three-dimensional rotation.
Then, using the Tonelli's theorem \cite{b4}, one obtains $m(\Omega_{t})=\int
m^{\prime}(\Omega_{t}(x))\mbox{d}x=0$ [$m^{\prime}(\cdot)$ is the Lebesgue
measure in $\mathbb{R}^{3}$]. Physically, this observation is intuitive. If
there is $\mathcal{K}_{0}=(\mathcal{P}_{0},\mathcal{P}_{0}^{\prime})\in
\Omega_{t}$, then $U_{\Lambda}^{\dag}(\mathcal{P}_{0})U_{\Lambda}%
(\mathcal{P}_{0}^{\prime})$ must have an eigenstate to be exactly $\rho$ (the
rest-frame reduced density matrix for spins). All such $(\mathcal{P}%
_{0},\mathcal{P}_{0}^{\prime})$ occupy only a low dimensional subset in
$\mathbb{R}^{6N}$ for any pure $\rho$ and nondegenerate $\Lambda$, thus
$m(\Omega_{t})=0$.

\textbf{Proof of the Theorem:} Because $m(\Omega_{g})>0$ while $m(\Omega
_{t})=0$, we have $\int_{\Omega_{g}}(\cdot)\mbox{d}\mathcal{K}\equiv
\int_{\Omega_{g}/\Omega_{t}}(\cdot)\mbox{d}\mathcal{K}$. Therefore
\begin{align*}
\operatorname{tr}(\rho^{\prime2})  & =\int_{\Omega_{g}}\mathcal{G}%
(\mathcal{K})\mathcal{T}(\mathcal{K})\mbox{d}\mathcal{K}=\int_{\Omega
_{g}/\Omega_{t}}\mathcal{G}(\mathcal{K})\mathcal{T}(\mathcal{K}%
)\mbox{d}\mathcal{K}\\
& <\int_{\Omega_{g}/\Omega_{t}}\mathcal{G}(\mathcal{K})\mbox{d}\mathcal{K}%
=\int_{\Omega_{g}}\mathcal{G}(\mathcal{K})\mbox{d}\mathcal{K}=1,
\end{align*}
where the inequality is due to that $\mathcal{T}(\mathcal{K})$ is
strictly
less than $1$ on $\Omega_{g}/\Omega_{t}$. Now that $\operatorname{tr}%
(\rho^{\prime2})<1$ immediately gives that $\rho^{\prime}$ is mixed.\hfill
$\blacksquare$

The fact that for localized states $m(\Omega_{g})>0$ is essential. Its
meanings can be further clarified by reviewing Eq. (\ref{eq 4}). If
$\operatorname{tr}(\rho^{\prime2})=1$, we must have $\mathcal{D}_{\lambda
}\langle\Delta\mathcal{Q}_{\lambda}^{2}\rangle=0$ for all $\lambda$. Supposing
there is $\lambda_{0}$ so that $\mathcal{D}_{\lambda_{0}}\neq0$, consequently
we have $\langle\Delta\mathcal{Q}_{\lambda_{0}}^{2}\rangle=0$. Due to that
$\mathcal{M}$ is orthogonal, there is $\sigma_{0}$ so that $\mathcal{M}%
_{\sigma_{0}\lambda_{0}}\neq0$. Supposing $\sigma_{0}$ corresponds the
position coordinate labelled by $\mu_{0}$ of particle $k_{0}$, we obtain that
$[\widehat{x}_{k_{0}\mu_{0}},\mathcal{\widehat{Q}}_{\lambda_{0}}%
]=i\hbar\mathcal{M}_{\sigma_{0}\lambda_{0}}$, which leads to that
$\langle\Delta x_{k_{0}\mu_{0}}^{2}\rangle\langle\Delta\mathcal{Q}%
_{\lambda_{0}}^{2}\rangle\geqslant\hbar^{2}\mathcal{M}_{\sigma_{0}\lambda_{0}%
}^{2}/4$. However because $\langle\Delta\mathcal{Q}_{\lambda_{0}}^{2}%
\rangle=0$ here, we see that $\langle\Delta x_{k_{0}\mu_{0}}^{2}%
\rangle\rightarrow\infty$. On the one hand, $\langle\Delta\mathcal{Q}%
_{\lambda_{0}}^{2}\rangle=0$ implies that $g(\mathcal{P})$ is an eigenstate of
$\mathcal{\widehat{Q}}_{\lambda_{0}}$ and consequently $m(\Omega_{g})=0$. On
the other hand, $\langle\Delta x_{k_{0}\mu_{0}}^{2}\rangle\rightarrow\infty$
implies that the (reduced) wave package of particle $k_{0}$ is nonlocalized.
Inversely, $m(\Omega_{g})>0$ guarantees that $g(\mathcal{P})$ is not an
eigenstate of any $\mathcal{\widehat{Q}}_{\lambda}$, so the state can be
localized because all $\langle\Delta x_{k\mu}^{2}\rangle$ can be finite.

\section{Discussion and Conclusion}

We would like to note that the present paper adopts the same
notion of spins as in Ref. \cite{r1,r2,r3}. Beside, there are
other possible notions (e.g. see Ref. \cite{r4}). It would be
interesting to see that when \textquotedblleft
spin\textquotedblright\ is defined with respect to projection of
Pauli-Lubanski's vector in a principal null direction of the
Lorentz transformation, the reduced density matrix for spins
viewed by the moving observer does not depolarize \cite{r4}.
However, since establishing a perfect shared reference frame
requires infinite communication even in non-relativistic
situations \cite{srf}, it might be extremely difficult to acquire
precise information about a Lorentz transformation. We argue that
in practical applications spin may be defined independently of the
particular Lorentz transformation that defines the relative motion
between the observers, and the transformation law of such spins
would then in general depend upon momentum. Indeed, our results
would hold for all such notions of spin, including those adopted
in Refs. \cite{r1,r2,r3}, but Ref. \cite{r4}.

In conclusion, states one can prepare in real experiments are
necessarily localized. Nonlocalized states, e.g. momentum
eigenstates and the singular ones presented in Ref. \cite{r3}, are
not practical in reality, although they are useful in theories. We
show that, in relativistic applications reduced spin state which
is pure in the rest frame unavoidably appears to be mixed whenever
viewed from moving reference frames. How much such boost-induced
decrease in purity is depends on how much the spatial wave package
is localized. The more the spatial wave package is localized, the
more the purity of the reduced spin state decreases when viewed
from moving frames. This observation may be important for
relativistic quantum information processing, particularly for
relativistic quantum communication.

Although our investigations are based on massive particles, the generalize to
massless cases, such as to photons \cite{r6}, should be analogous. This may be
of interest since most of current experiments in quantum communication are
based on photons \cite{b5}. Another interesting problem might be to determine
how our results generalize to accelerated frames \cite{r7}.

\begin{acknowledgments}
We thank S.X. Yu and Z.B. Chen for useful discussion. This work
was supported by the Nature Science Foundation of China (Grant No.
10075041), the National Fundamental Research Program (Grant No.
2001CB309300), and the ASTAR Grant No. 012-104-0040 \&
R-144-000-071-305.
\end{acknowledgments}


\begin{thebibliography}{99}                                                                                               %
\bibitem {b1}R.C. Tolman, \textit{Relativity, Thermodynamics, and Cosmology}
(Oxford University Press, Oxford, 1934); R.M. Wald, \textit{Quantum Field
Theory in Curved Spacetime and Black Hole Thermodynamics} (University of
Chicago Press, Chicago, 1994).

\bibitem {r0}P.T. Landberg and G.E.A. Matsas, Phys. Lett. A \textbf{223}, 401
(1996); P.J.B. Peebles and D.T. Wilkinson, Phys. Rev. \textbf{174}, 2168 (1968).

\bibitem {r1}A. Peres \textit{et al.}, \prl \textbf{88}, 230402 (2002).

\bibitem {r2}R.M. Gingrich and C. Adami, \prl \textbf{89}, 270402 (2002).

\bibitem {r3}H. Li and J. Du, \pra \textbf{68}, 022108 (2003).

\bibitem {r4}M. Czachor and M. Wilczewski, \pra \textbf{68}, 010302(R) (2003).

\bibitem {r5}M. Czachor, \pra \textbf{55}, 72 (1996); J. Rembieli\'{n}ski and
K.A. Smoli\'{n}ski, \pra \textbf{66}, 052114 (2002); P.M. Alsing and G.J.
Milburn, Quantum Inf. and Comput. \textbf{2}, 487 (2002); D. Ahn \textit{et
al.}, \pra \textbf{67}, 012103 (2003).

\bibitem {r6}R.M. Gingrich et al., \pra\textbf{68}, 042102 (2003); A. Peres
and D.R. Terno, J. Mod. Opt. \textbf{50}, 1165 (2003).

\bibitem {r7}P.M. Alsing and G.J. Milburn, \prl\textbf{91}, 180404 (2003).

\bibitem {b5}\textit{The Physics of Quantum Information}, edited by D.
Bouwmeester, A. Ekert, and A. Zeilinger (Springer, New York, 2000).

\bibitem {ft1}In Ref. \cite{r3}, it is shown that a class of states of two
spin half particles, other than momentum eigenstates, do have covariant
reduced density matrix for spins. However, it can be proved that for such
states the reduced state for a single particle is the convex combination of
momentum eigenstates. Taking the momentum wave function $g(\mathbf{p}%
_{1},\mathbf{p}_{2})=[f(\mathbf{p}_{1})\delta^{3}(\mathbf{p}_{1}%
-\mathbf{p}_{2})]^{1/2}$ (see Ref. \cite{r3}) as an example, and noting that
$g(\mathbf{p}_{1},\mathbf{p}_{2})=g(\mathbf{p}_{2},\mathbf{p}_{1})$ and
$\delta(x)^{1/2}\equiv\delta(x)/[\delta(0)^{1/2}]$, the reduced state of
either particle can be written as a density matrix with elements being
$\varrho(\mathbf{p},\mathbf{p}^{\prime})=\int g^{\ast}(\mathbf{p}%
,\mathbf{q})g(\mathbf{p}^{\prime},\mathbf{q})\mbox{d}\mathbf{q}=|f(\mathbf{p}%
)|\delta^{3}(\mathbf{p}-\mathbf{p}^{\prime})/[\delta(0)^{3}]=|f(\mathbf{p}%
)|\delta_{\mathbf{p},\mathbf{p}^{\prime}}$. The similar holds for other
singular states presented in Ref. \cite{r3}.

\bibitem {b3}S. Weinberg, \textit{The Quantum Theory of Fields} (Cambridge
University Press, Cambridge, England, 1996).\-

\bibitem {b4}S.K. Berberian, \textit{Measure and Integration} (The Macmillan
Company, New York, 1962).

\bibitem {srf}A. Peres and P. F. Scudo, \prl\textbf{86}, 4160 (2001),
\prl\textbf{87}, 167901 (2001); E. Bagan et al., \prl\textbf{87}, 257903 (2001).
\end{thebibliography}
\end{document}